\documentclass[]{article}
\usepackage{style}

\usepackage[utf8]{inputenc} 
\usepackage[T1]{fontenc}    
\usepackage{hyperref}       
\usepackage{url}            
\usepackage{booktabs}       
\usepackage{amsfonts}       
\usepackage{nicefrac}       
\usepackage{microtype}      
\usepackage{lipsum}
\usepackage{fancyhdr}       
\usepackage{graphicx}       
\usepackage{multicol}
\usepackage{longtable}
\usepackage{tabularx}
\usepackage{float}
\usepackage{caption}
\usepackage{setspace} 
\usepackage{placeins}
\usepackage{array}
\usepackage{tabularx}  
\usepackage{booktabs}  
\usepackage[
backend=biber,
style=ieee,
citestyle=numeric-comp
]{biblatex}

\usepackage[utf8]{inputenc}
\usepackage[T1]{fontenc}

\DeclareUnicodeCharacter{1EC7}{e}

\AtEveryBibitem{\clearfield{address}}
\AtEveryBibitem{\clearlist{address}}

\AtEveryBibitem{\clearfield{place}}
\AtEveryBibitem{\clearlist{place}}

\AtEveryBibitem{\clearfield{issn}}
\AtEveryBibitem{\clearlist{issn}}

\AtEveryBibitem{\clearfield{series}}
\AtEveryBibitem{\clearlist{series}}

\AtEveryBibitem{\clearfield{month}}
\AtEveryBibitem{\clearlist{month}}

\AtEveryBibitem{\clearfield{isbn}}
\AtEveryBibitem{\clearlist{isbn}}

\AtEveryBibitem{\clearfield{language}}
\AtEveryBibitem{\clearlist{language}}

\AtEveryBibitem{\clearfield{doi}}
\AtEveryBibitem{\clearlist{doi}}

\AtEveryBibitem{%
  \ifentrytype{misc}
    {%
    }
    {%
      \clearfield{url}
      \clearlist{url}
    }%
}

\AtEveryBibitem{\clearfield{extra}}
\AtEveryBibitem{\clearlist{extra}}

\AtEveryBibitem{\clearfield{note}}
\AtEveryBibitem{\clearlist{note}}

\AtEveryBibitem{%
  \ifentrytype{online}
    {}
    {\clearfield{urlyear}\clearfield{urlmonth}\clearfield{urlday}}}

\graphicspath{{media/}} 

\title{Large Language Models Can Infer Personality from Free-Form User Interactions
}

\author{
Heinrich Peters            \\
Columbia University\\
\texttt{\ hp2500@columbia.edu}\\
\And
Moran Cerf\\
Columbia University \\
\texttt{\ mc5424@columbia.edu }\\
\And
Sandra C. Matz\\
Columbia University\\
\texttt{\ sm4409@columbia.edu}\\
}

\begin{document}

\maketitle


\vspace{3cm}

\begin{abstract}
This study investigates the capacity of Large Language Models (LLMs) to infer the Big Five personality traits from free-form user interactions. The results demonstrate that a chatbot powered by GPT-4 can infer personality with moderate accuracy, outperforming previous approaches drawing inferences from static text content. The accuracy of inferences varied across different conversational settings. Performance was highest when the chatbot was prompted to elicit personality-relevant information from users (mean r=.443, range=[.245, .640]), followed by a condition placing greater emphasis on naturalistic interaction (mean r=.218, range=[.066, .373]). Notably, the direct focus on personality assessment did not result in a less positive user experience, with participants reporting the interactions to be equally natural, pleasant, engaging, and humanlike across both conditions. A chatbot mimicking ChatGPT’s default behavior of acting as a helpful assistant led to markedly inferior personality inferences and lower user experience ratings but still captured psychologically meaningful information for some of the personality traits (mean r=.117, range=[-.004, .209]). Preliminary analyses suggest that the accuracy of personality inferences varies only marginally across different socio-demographic subgroups. Our results highlight the potential of LLMs for psychological profiling based on conversational interactions. We discuss practical implications and ethical challenges associated with these findings.
\end{abstract}
\vspace{1cm}




\newpage
\section{Introduction}
Large Language Models (LLMs) have emerged as an important technological innovation with far-reaching implications for research and practice. Models like OpenAI’s GPT-4 \cite{openai_gpt-4_2023}, Anthropic’s Claude \cite{anthropic_model_2023}, Meta's Llama \cite{touvron_llama_2023-1} or Google’s Gemini \cite{gemini_team_gemini_2023} show a remarkable ability to understand and generate human-like text. In a striking departure from previous modeling approaches, the capabilities of LLMs tend to generalize well to previously unseen scenarios, contexts, and tasks \cite{brown_language_2020, radford_language_2019}. This includes their ability to represent complex aspects of human psychology.  

As past work has shown, LLMs can solve theory of mind tasks, indicating their ability to impute mental states to other entities \cite{kosinski_theory_2023}. Similarly, it has been demonstrated that LLMs exhibit behavioral patterns that map onto psychometric inventories originally developed to measure individual differences in humans. While past research has raised concerns about the validity of such questionnaires when applied to LLMs \cite{kovac_large_2023, dorner_personality_2023}, commonly used LLMs seem to differ systematically in their reactions to items from standard inventories assessing non-cognitive psychological characteristics such as personality, values, moral norms, and diversity beliefs \cite{pellert_ai_2024}. Inversely, LLMs are able to produce behavior that is aligned with human traits when prompted to do so. For example, a model that is instructed to adopt an extraverted persona would score high on that trait when presented with a personality questionnaire \cite{jiang_evaluating_2023, serapio-garcia_personality_2023}.

In addition to mimicking aspects of human psychology, LLMs can also decode people’s psychological traits from various forms of user-generated text. For example, past work has shown that LLMs can infer personality from essays using chain-of-thought reasoning based on items from personality inventories \cite{yang_psycot_2023}. Similarly, past work has examined the ability of LLMs to infer personality from transcripts of asynchronous video interviews \cite{zhang_can_2024}, as well as social media posts \cite{peters_large_2023}. In many cases, the predictions made by LLMs can compete with those made by supervised machine learning models. Yet, given the fact that LLMs do not require researchers to collect their own data and train a separate model for every psychological dimension of interest, they offer an unprecedented opportunity to investigate and apply psychological traits at scale and outside the lab \cite{peters_large_2023}.

As the outlined examples show, most existing research evaluating the capabilities of LLMs to make inferences about psychological traits has focused on their capacity to interpret static written content using relatively simple prompting strategies. Yet, many real-world applications of LLMs involve dynamic interactions between AI chatbots and users. Compared to static data sources, these conversational interactions provide a more naturalistic setting that closely mirrors everyday human interactions and can have similar psychological effects on users \cite{ho_psychological_2018}. Conversations also offer unique opportunities for personality judgments as they allow AI chatbots to elicit personality-relevant information from users. Relatedly, conversations might reveal more subtle aspects of an individual's psychological makeup, such as their ability to adapt to new information, manage conversational flow, and respond to emotional cues.

Building on the unique opportunities afforded by naturalistic user interactions with LLM-based chatbots, this paper examines the ability of LLMs to infer personality traits from free-flowing conversations. In doing so, it addresses three related research questions. First, we examine whether LLM-based chatbots can infer people’s Big Five personality traits under a series of conditions that provide different instructions to the chatbots and users. Specifically, we prompted the LLM chatbot to either (i) assess their users' personality, (ii) have a naturalistic conversation, or (iii) act as a “helpful assistant,” a commonly used default setup in popular chatbots like ChatGPT. At the same time, we instructed users to (i) have a naturalistic conversation or (ii) freely use the chatbot according to their own preferences (e.g., for question answering). Second, we analyze potential trade-offs between the accuracy of personality inferences and user experience across conditions. For example, prompting the model to elicit personality-relevant information from users might negatively impact their appraisal of the conversation. Finally, we analyze potential biases in LLM-based personality inferences with regard to several important demographic characteristics, such as gender, age, race, education, and socioeconomic status.

\section{Method}
\subsection{Sample and Data Collection}
We recruited 600 US participants through Prolific Academic in February 2024. Out of the recruited participants, 32 dropped out during data collection, and two were removed for failing to pass attention checks. In the final sample of 566 participants, 50.2\% of participants identified as female, 62.7\% identified as White, and 55.7\% had obtained at least a college degree. The average age was 37.67 years (SD=13.01). All participants gave informed consent. The research was covered under the Columbia University IRB (Protocol \#AAAV0800).

\subsection{Research Design and Procedure}
We employed a 3x2 factorial between-subjects design to test ChatGPT’s ability to predict psychological traits across different interaction modes (i.e., prompts and instructions given to ChatGPT and participants). The different chatbots were prompted to (i) evaluate the users’ big five personality traits in a naturalistic conversation (assessment condition), (ii) get to know the user in a naturalistic conversation (acquaintance condition), or (iii) act as a helpful assistant (assistant condition). Similarly, participants were instructed to (i) have a naturalistic conversation (conversation condition) or (ii) interact with the chatbot however they wanted (unconstrained use condition). Participants were randomly assigned to one of the resulting six conditions.

After participants had provided informed consent, they were directed to a web app where they interacted with a chatbot built on top of the ChatGPT API. As is common in AI-powered chatbots, the conversation was structured such that participants and the chatbots took turns of one message each. Participants were made aware that they would interact with an AI agent, but they did not receive any information about the purpose of the study. The interaction lasted for a predetermined length of 15 turns per participant, during which participants could freely interact with the chatbot, asking questions and answering questions about themselves. After 15 turns, participants received a confirmation code and were directed to an online questionnaire, where they were presented with a brief user experience survey and a personality questionnaire (BFI-2) \cite{soto_next_2017}. Moreover, participants were prompted to provide basic socio-demographic information (age, gender, ethnicity, income, and education). The median completion time was 19 minutes and 17 seconds.

\begin{table}[ht]
\small
\centering
\renewcommand{\arraystretch}{1.5}
\begin{tabularx}{\textwidth}{>{\raggedright\arraybackslash}p{2.5cm} >{\raggedright\arraybackslash}p{2.5cm} >{\raggedright\arraybackslash}X}
\toprule
Condition Type & Condition Name & System Prompt / Instruction \\ 
\midrule
Bot Conditions & Assessment & Have a natural conversation with the user, trying to get to know them and rate their Big-5 personality traits Openness, Conscientiousness, Extraversion, Agreeableness, Neuroticism. Make sure it does not feel like an assessment and under no circumstances let them know you are evaluating them. Do not provide feedback on their personality during the conversation. Keep the conversation going for as long as possible and continue to collect useful information. Be concise in your questions and responses. \\ 
\cmidrule(l){2-3}
 & Acquaintance & Have a natural conversation with the user, trying to get to know them and get a sense of their character. Imagine you meet them for the first time. Keep the conversation going for as long as possible and continue to collect useful information. Be concise in your questions and responses. \\ 
\cmidrule(l){2-3}
 & Assistant & You are a helpful assistant. \\ 
\midrule
User Conditions & Conversation & In the first part of the study, you will engage in a short conversation with a chatbot. The idea is to have a natural conversation - the same way you would have with a person you've just met. There are no right or wrong answers, but please make sure that you engage in a real and meaningful exchange (we will otherwise not be able to use your responses). \\ 
\cmidrule(l){2-3}
 & Unconstrained Use & In the first part of the study, you will engage with a chatbot. The idea is to use it the way you would typically use a bot like ChatGPT. You can ask questions, make it solve tasks for you, or simply have a conversation with it. There are no right or wrong questions, but please make sure that you engage in a real and meaningful user experience (we will otherwise not be able to use your responses). \\ 
\bottomrule
\end{tabularx}
\vspace{6pt}
\caption{Overview of system prompts and user instruction across bot and user conditions. Bot conditions manipulate the behavior of the LLM through prompting. User conditions manipulate the behavior of the user by instructing them to interact with the bot in a particular way.}
\end{table}

\subsection{Materials and Measures}
The chatbot user interface was set up as a custom web application with a chat window, running on Google AppEngine. We constructed the web application on top of the ChatGPT API using the GPT-4 \cite{openai_gpt-4_2023} model version gpt-4-0613. All messages and message metadata were saved to a Firestore NoSQL database in real time. In order to generate inferred personality scores, each conversation was parsed into a transcript that included the messages from the chatbot and the user, marked as “GPT: …” and “User: …” and then fed back to the model. Inferred personality scores were obtained by making three additional API calls, each prompting the GPT-4 model to return personality scores based on the transcript of the conversation. The scores used in all analyses are the means across three rounds of scoring. ChatGPT system prompts and user interactions for the different conditions can be found in Table 1. Assessment prompts can be found in SI \ref{inference_prompt}.

The questionnaire participants completed after their interaction with the chatbot consisted of six questions aiming to evaluate different aspects of the user experience: 1) “The conversation was natural.” 2) “The conversation was pleasant.” 3) “The conversation was engaging.” 4) “My conversation partner asked good questions.” 5) “My conversation partner gave good answers.” 6) “My conversation partner was humanlike.” Responses were coded on a 7-point Likert scale. While the items were not intended apriori to measure a single construct, the high internal consistency of Alpha=0.898 (all Alpha>0.843 when computed separately within conditions) suggests that an aggregate score could be interpreted as a positive versus negative appraisal of the overall user experience. 

In addition to the user experience items, self-reported personality scores were captured using the BFI-2 questionnaire \cite{soto_next_2017}, a widely used measurement with excellent psychometric properties. The questionnaire consists of 60 items with a 5-point Likert scale answer format, ranging from 1 = strongly disagree to 5 = strongly agree. Socio-demographic characteristics were captured using single-choice and multiple-choice menus for gender, age, education, race/ethnicity, and income. The full list of items, including all response options, can be found in SI \ref{questionnaires}.

\section{Results}
\subsection{Can LLMs Infer Personality Traits from Free-Form User Interactions?}
We tested ChatGPT's ability to infer users' Big Five personality traits by computing Pearson’s correlation coefficients between self-reported and inferred scores separately for each condition. We used one-sided significance tests to test whether the correlations were greater than zero
 
As expected, the accuracy of personality inferences was the highest in the assessment conditions where the chatbot was explicitly prompted to collect information about the user’s big five personality traits, with correlations ranging from r=.326 to r=.590 (mean r=.438) when users were tasked to have a naturalistic conversation with the chatbot. When users interacted with the chatbot in an unconstrained fashion, the correlations ranged from r=.245 to r=.640 for (mean r=.448). All correlations were significantly greater than zero and there was no significant difference in correlations across both user conditions.

The accuracy of inferred scores in the acquaintance conditions where the chatbot was instructed to get to know the user was overall lower than in the assessment condition. When users were tasked to have a naturalistic conversation with the chatbot, the correlations between self-reported and inferred trait scores ranged from r=.166 to r=.373 (mean r=.248). When users interacted with the chatbot in an unconstrained fashion, the correlations ranged from r=.066 to r=.317 (mean r=.188). The majority of correlations were significantly greater than zero. The correlations did not significantly differ across the two user conditions. 

Finally, the accuracy of inferred scores was the lowest in the assistant condition, where the chatbot did not ask personal questions and simply reacted to user prompts. Yet, the correlations between self-reported and inferred trait scores were predominantly positive and significantly greater than zero for several traits, highlighting that even users’ regular day-to-day interactions with bots like ChatGPT can contain psychologically meaningful information. When users were tasked to have a naturalistic conversation with the chatbot, the correlations between self-reported and inferred trait scores ranged from r=-.004 to r=.209 for (mean r=.126). When users interacted with the chatbot in an unconstrained fashion, the correlations ranged from r=.018 to r=.193 (mean r=.109). As before, the correlations did not differ significantly across both user conditions. 

The outlined findings suggest that the prompted behavior of the chatbot greatly impacts the quality of personality inferences by eliciting more or less relevant information. The user instructions (conversation vs unconstrained use), on the other hand, did not appear to have a strong impact on the accuracy of inferences. For a detailed presentation of the results, please refer to Figure 1 or SI \ref{accuracy_across_conditions}.

\begin{figure}[t]
  \centering
  \includegraphics[width=1\textwidth]{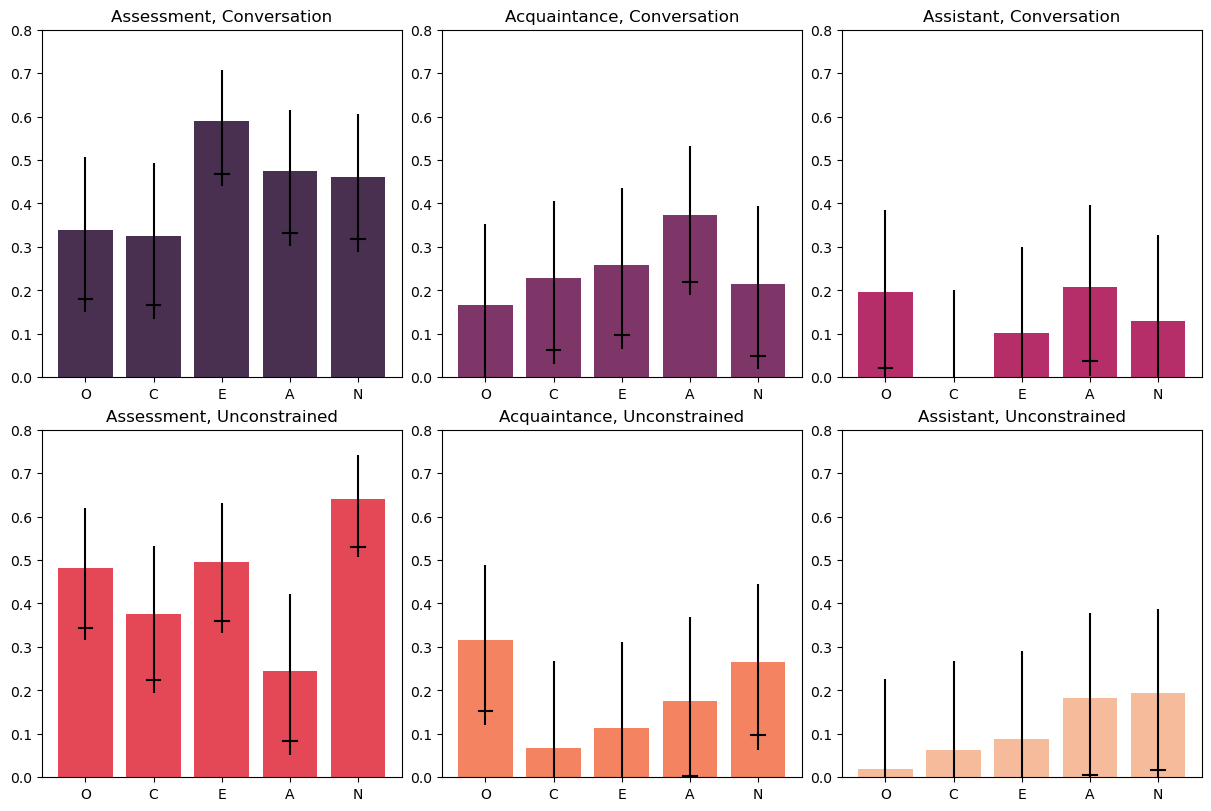}
  \caption
  {Correlations between inferred and self-reported personality trait scores across conditions. ChatGPT conditions (assessment, acquaintance, assistant) are shown in different columns. User conditions (conversation, unconstrained) are shown across rows. The vertical black lines represent two-tailed 95\% confidence intervals. The horizontal black lines represent one-tailed lower 95\% confidence intervals.}
  \label{fig:correlations}
\end{figure}

\subsection{How Do Different Interaction Modes Affect User Experience?}
We examined the potential trade-offs between inference accuracy and user experience across conditions by comparing participants' ratings on six dimensions (i.e., whether the interaction was natural, pleasant, and engaging, the perceived quality of the chatbot’s questions and answers, and whether the chatbot was perceived as humanlike). Participants rated their interaction as the most natural, pleasant, and engaging in the acquaintance condition and as the least natural and humanlike in the assistant condition. Importantly, however, there was no systematic pattern of significant differences in participants’ experience between the assessment and acquaintance conditions. This suggests that the chatbot’s direct focus on personality-relevant topics – and the resulting boost in accuracy – does not necessarily come at the cost of a less positive user experience. For a visual presentation of the mean user experience ratings across conditions, please refer to Figure 2. Detailed statistics can be found in SI \ref{user_experience}

\begin{figure}[t]
  \centering
  \includegraphics[width=1\textwidth]{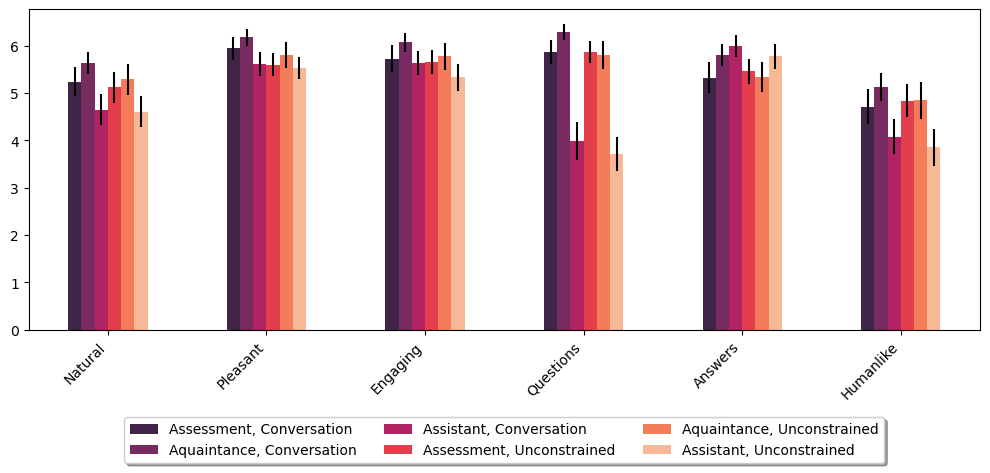}
  \caption
  {User experience ratings across items (The conversation was natural; The conversation was pleasant; The conversation was engaging; My conversation partner asked good questions; My conversation partner gave good answers; My conversation partner was humanlike) and conditions. The vertical black lines represent two-sided 95\% confidence intervals.}
  \label{fig:ux}
\end{figure}

\subsection{Does the Quality of LLM-based Personality Inferences Vary Across Demographic Groups?}

In order to examine potential biases in personality inferences, we analyzed the differences in the residuals of inferred and self-reported scores, as well as differences in the correlations between inferred and self-reported scores across demographic groups. Residuals were computed by subtracting the self-reported scores from the inferred scores, such that positive values represent an overestimation in inferred scores and negative values represent underestimations. Correlations were computed separately within each demographic group such that a lower correlation coefficient in one group would indicate a lower level of accuracy compared to the other group. To achieve sufficient coverage within subgroups, we pooled the sample across conditions and dichotomized each of the five sociodemographic variables of interest: age (median split at 36 years), race (non-white, white), education (no college education, college education), and income (less than \$50k, more than \$50k). Distribution statistics and group sizes are reported in SI \ref{demographics}.
 
Figure 3 reports the residuals and correlations across subgroups for five different socio-demographic variables. As the residual graphs on the left suggest, ChatGPT tended to underestimate participants’ personality traits across the board. However, while there were a few instances with significant differences across subgroups (e.g., gender differences for Neuroticism, or age differences for Conscientiousness), the overall trend seems to suggest a relatively unbiased prediction accuracy across the five socio-demographic categories and personality traits. The same pattern was found for the within-group correlations, with only two significant differences for income and Conscientiousness and Agreeableness. Detailed statistics can be found in SI \ref{group_differences}.

\begin{figure}[h]
  \centering
  \includegraphics[width=1\textwidth]{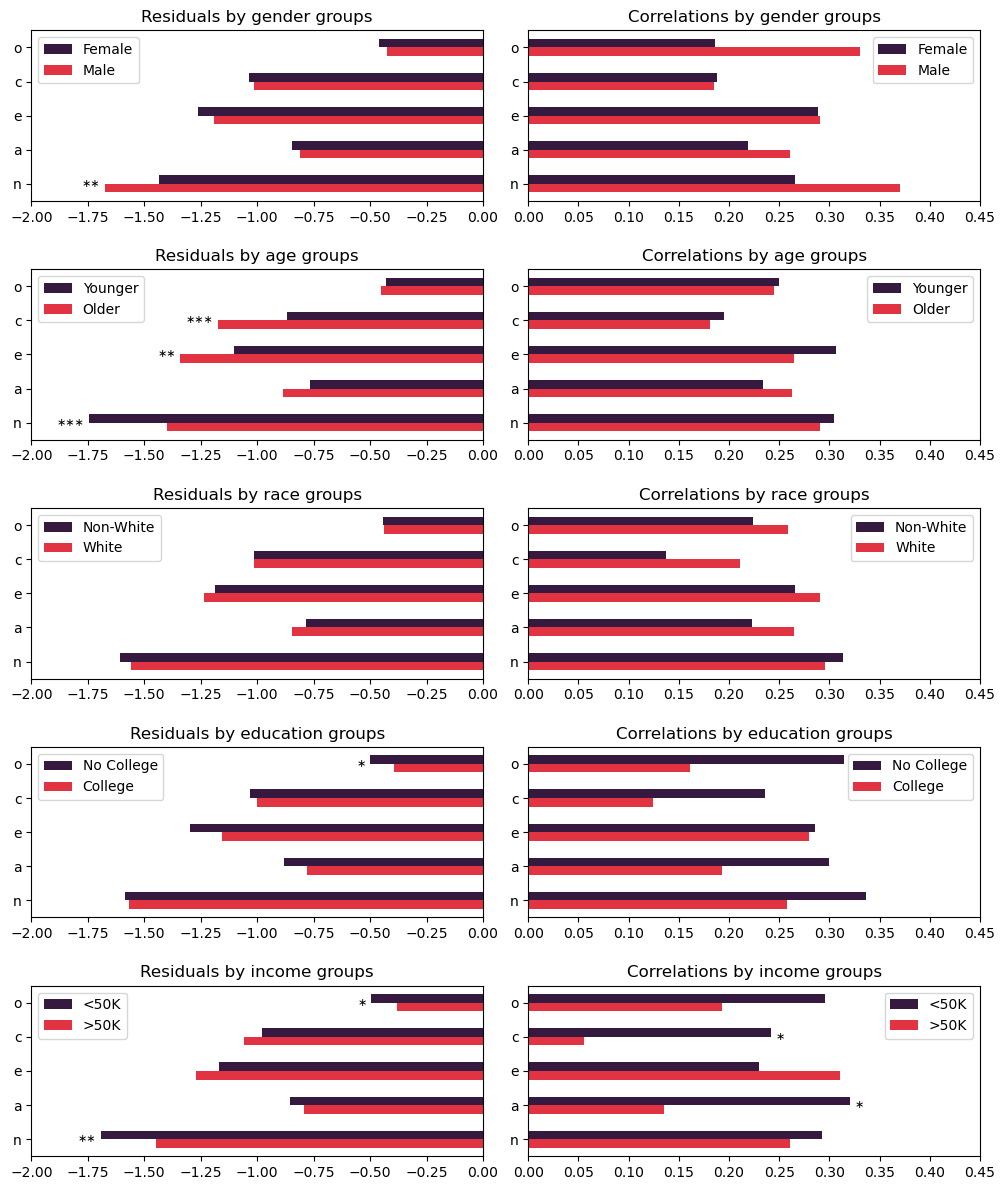}
  \caption
  {Group differences in residuals (left) and correlations (right) across demographic groups. Residuals were computed as the difference between inferred and self-reported personality scores, such that a negative value indicates a negative bias in inferred scores. A larger bar indicates larger residuals for the specified group. Correlations were computed as Pearson’s correlation coefficients between inferred and self-reported personality scores within each group. A larger bar indicates higher accuracy for the specified group.}
  \label{fig:subgroups}
\end{figure}

\FloatBarrier

\section{Discussion}
\subsection{Interpretation of Results}
The present study investigates the ability of LLMs to infer the big five personality traits from free-form user interactions. Highlighting the potential of conversational approaches, the personality inferences were more accurate than those reported in previous work based on static text in combination with both supervised learning \cite{grunenberg_machine_2024, park_automatic_2015, schwartz_personality_2013} and LLM-powered approaches \cite{peters_large_2023}. Notably, both the acquaintance and assistant conditions provided meaningful psychological information for better-than-chance performance. While the accuracy in the acquaintance condition was still on par with results from the prior literature for several traits, the accuracy in the assistant condition was expectedly lower. The overall pattern of results suggests that LLMs possess accurate representations of personality constructs and are able to elicit personality-relevant information by asking pointed questions.

We did not observe systematic differences in user experience between the assessment and acquaintance conditions, suggesting that there are no stark trade-offs between accuracy and user experience regarding personality inferences from free-flowing conversations. Practitioners and researchers can hence implement a prompting strategy targeting personality-related information without the concern of alienating users. Both the assessment and the acquaintance conditions were judged more favorably than the assistant conditions, which only stood out with regard to users’ assessment of the quality of answers provided by the chatbot. This preference suggests that a more naturalistic interaction style may enhance users’ experience with LLMs and is aligned with prior work showing reciprocity effects in user interactions with chatbots \cite{harmsen_eliciting_2023}.

The comparison of accuracies across subgroups suggests that there is little heterogeneity across different socio-demographic variables, both with regard to residuals and within-group correlations. The few exceptions could be explained by a number of different factors. First, the training data for LLMs may contain biases that skew toward the experiences and expressions typical of certain demographic groups, particularly those that are overrepresented in the data \cite{bolukbasi_man_2016, wan_biasasker_2023, abdurahman_perils_2023, durmus_towards_2024, santurkar_whose_2023, atari_which_2023, rathje_gpt_2023}. Additionally, language use and cultural factors might influence how personality traits are expressed and perceived \cite{church_current_2010, church_personality_2016}, potentially affecting the model's ability to accurately interpret and score these traits across diverse groups. However, given the small sample size and the resulting need to dichotomize the demographic variables, we consider these findings preliminary and encourage future research.

Taken together, our results highlight the potential of LLMs for psychological profiling based on conversational interactions and that interaction modes greatly influence the efficacy of LLM-based personality assessments.

\FloatBarrier

\subsection{Limitations and Directions for Future Work}
There are several limitations and areas for future research that warrant consideration. First, despite the conversations unfolding across multiple turns, the study investigates a zero-shot learning scenario with very simple prompts. It is likely that a more sophisticated prompting strategy (e.g., chain-of-thought prompting) \cite{yang_psycot_2023} or additional fine-tuning \cite{karra_estimating_2023} could yield more accurate personality inferences. Our results, therefore, represent a lower bound of inference performance. Additionally, as the interactions were relatively short, it stands to reason that longer or repeated interactions could provide a richer data set, potentially leading to better inferences \cite{peters_large_2023, youyou_computer-based_2015}. 

Second, the experimental setting may not fully capture the quality of intrinsically motivated user interactions that occur in more naturalistic settings. For instance, in the assistant condition, users engaging with the chatbot out of genuine need or interest – rather than being asked to do so – might reveal more about their daily circumstances, possibly leading to improved personality inferences. Future research should examine this hypothesis by utilizing user data from non-experimental settings.

Third, our study was conducted exclusively with ChatGPT, one of many LLMs available to the public. As there is a general convergence in the capabilities of different LLMs \cite[see][]{anthropic_model_2023, openai_gpt-4_2023, gemini_team_gemini_2023}, we expect that our results would translate across various other models. However, the precise extent of generalizability of our findings to a broader range of models remains a question for future research.

\subsection{Implications}
Our study has implications for a wide range of domains, including social science research, personalization, privacy, and AI ethics. On the one hand, the ability of LLMs to infer personality from free-form user interactions offers promising avenues for psychological assessment at scale. While the assessment of individual differences had previously been limited by the need to ask people stylized questions as part of personality tests \cite[e.g.,][]{mccrae_validation_1987, soto_next_2017} or access historical records of their digital footprints \cite{kosinski_facebook_2015, kosinski_private_2013, park_automatic_2015, youyou_computer-based_2015} the ability to infer personality through more dynamic, conversational approaches could help scale automated assessments and expand the study of individual differences to larger and more diverse samples. LLMs can, therefore, contribute to novel insights into human behavior and psychology \cite[see][]{demszky_using_2023, ziems_can_2023}.

Beyond academic research, the ability to scale psychometric assessments via LLM-based chatbots makes individual differences a viable contestant for large-scale personalization across a broad variety of domains, creating both opportunities and risks. As prior work has suggested, LLMs are not only able to make accurate personality inferences but also to craft persuasive messages tailored to people’s personality traits \cite{matz_potential_2024}. Given our results, it would be possible for artificial agents to interact with internet users to unobtrusively collect information about their psychological makeup and then use this information for targeted advertisements, customized product recommendations, or political campaigns. While such automated content recommendation and creation could facilitate more engaging and personalized user experiences online, they could also influence behavior in unprecedented and potentially harmful ways \cite{matz_privacy_2020}.
The tension between opportunities and the potential for misuse calls for a balanced approach to the deployment and regulation of AI technologies. We recommend clear regulations regarding the identification of artificial agents when they interact with humans and a transparent disclosure of the inferences made by the AI as a result of these interactions.

\subsection{Conclusion}
This study demonstrates the potential of LLMs to infer the Big Five personality traits from free-form user interactions, surpassing the performance of previous supervised learning approaches based on static text data. The findings highlight the effect of conversational context on the accuracy of LLM-based personality inferences while showing that user experience remains highly positive even when LLMs are specifically prompted to solicit personality-relevant information. Together, these insights open up new avenues for social science research and the application of personalization at scale. However, they also raise ethical and privacy concerns regarding the use of LLMs in psychological profiling and assessment.

\newpage

\section*{Ethics Approval}
The study was approved by the Columbia University IRB (Protocol \#AAAV0800). All methods were carried out in accordance with relevant guidelines and regulations.

\section*{Availability of Data and Materials}
All data and materials are available on this project's OSF page https://osf.io/edn3g/.

\section*{Competing Interests}
The authors declare no potential conflicts of interest.

\section*{Author Contributions}
HP: Conceptualization, Methodology, Software, Formal analysis, Investigation, Data curation, Writing - Original Draft;
MC: Funding acquisition; 
SCM: Conceptualization, Methodology, Writing - Original Draft, Funding acquisition

\section*{Acknowledgments}
The research was supported by the Digital Future Initiative at Columbia Business School.

\newpage
\printbibliography

\newpage
\section*{Supplementary Information}
\appendix

\section{Inference Prompt}
\label{inference_prompt}
\{"role": "system", "content": "You are a helpful assistant."\} \\
\{"role": "user", "content": "Rate the personality of the person called "user" on the Big Five personality dimensions. Pay attention to how people's personalities might be reflected in the way they respond to questions and what they share about themselves. Provide your response on a scale from 1 to 5 for the traits Openness, Conscientiousness, Extraversion, Agreeableness, and Neuroticism. Provide only the numbers.”\}\\
\{role": "assistant", "content": "Please provide your input."\}\\
\{role": "user", "content": "Please rate this text: " + input\_string + "This is the end of the conversation."\}\\
\{role": "system", "content": "Do not respond to anything that came up in the conversation. Just rate the user's personality. Remember to provide your response on a scale from 1 to 5 for the traits Openness, Conscientiousness, Extraversion, Agreeableness, and Neuroticism. Pay attention to how people's personalities might be reflected in the way they respond to questions and what they share about themselves. Provide only the numbers. Only provide integers, no decimals.”\}\\

\section{Questionnaires}
\label{questionnaires}
\subsection{User Experience Questionnaire}
\begin{itemize}
    \item The conversation was natural.
    \item The conversation was pleasant.
    \item The conversation was engaging.
    \item My conversation partner asked good questions.
    \item My conversation partner gave good answers.
    \item My conversation partner was humanlike.
\end{itemize}

\subsection{Demographic Questionnaire}
\begin{itemize}
  \item What is your gender? (single choice: male; female; other)
  \item What year were you born? (single choice drop-down menu)
  \item What is the highest level of education you have completed? (single choice: Some high school or less; High school diploma or GED; Some college, but no degree; Associate or technical degree; Bachelor’s degree; Graduate or professional degree)
  \item Choose one or more races that you consider yourself to be (multiple choice: White or Caucasian; Black or African American; American Indian/Native American or Alaska Native; Asian; Native Hawaiian or Other Pacific Islander; Other)
  \item What is your employment status? (single choice: Employed; Self-employed; Unemployed; Student; Retired; Other)
  \item Annual personal income before tax (this includes all earnings, social security income, public assistance, \& retirement income) (single choice: Less than \$10,000; \$10,000 to \$14,999; \$15,000 to \$24,999; \$25,000 to \$34,999; \$35,000 to \$49,999; \$50,000 to \$74,999; \$75,000 to \$99,999; \$100,000 to \$149,999; \$150,000 to \$199,999; \$200,000 or more)
\end{itemize}

\newpage
\section{Inference Accuracy Across Conditions}
\label{accuracy_across_conditions}
\FloatBarrier
\begin{table}[h]
    \begin{tabular*}{\textwidth}{@{\extracolsep{\fill}}lrrrrrrrr}
\toprule
model & bot condition & user condition & trait & r & ci\_l & ci\_u & ci\_l\_os & p \\
\midrule
gpt-4 & Assessment & Conversation & o & 0.340 & 0.149 & 0.506 & 0.181 & 0.000 \\
gpt-4 & Assessment & Conversation & c & 0.326 & 0.134 & 0.494 & 0.166 & 0.001 \\
gpt-4 & Assessment & Conversation & e & 0.590 & 0.441 & 0.707 & 0.467 & 0.000 \\
gpt-4 & Assessment & Conversation & a & 0.474 & 0.303 & 0.616 & 0.332 & 0.000 \\
gpt-4 & Assessment & Conversation & n & 0.462 & 0.288 & 0.606 & 0.318 & 0.000 \\
gpt-4 & Acquaintance & Conversation & o & 0.166 & -0.033 & 0.352 & -0.001 & 0.051 \\
gpt-4 & Acquaintance & Conversation & c & 0.228 & 0.032 & 0.407 & 0.064 & 0.012 \\
gpt-4 & Acquaintance & Conversation & e & 0.259 & 0.065 & 0.434 & 0.097 & 0.005 \\
gpt-4 & Acquaintance & Conversation & a & 0.373 & 0.189 & 0.531 & 0.220 & 0.000 \\
gpt-4 & Acquaintance & Conversation & n & 0.215 & 0.018 & 0.395 & 0.050 & 0.016 \\
gpt-4 & Assistant & Conversation & o & 0.195 & -0.011 & 0.386 & 0.022 & 0.032 \\
gpt-4 & Assistant & Conversation & c & -0.004 & -0.210 & 0.202 & -0.178 & 0.516 \\
gpt-4 & Assistant & Conversation & e & 0.101 & -0.107 & 0.301 & -0.074 & 0.170 \\
gpt-4 & Assistant & Conversation & a & 0.209 & 0.003 & 0.398 & 0.036 & 0.024 \\
gpt-4 & Assistant & Conversation & n & 0.130 & -0.078 & 0.327 & -0.044 & 0.110 \\
gpt-4 & Assessment & Unconstrained & o & 0.482 & 0.315 & 0.619 & 0.344 & 0.000 \\
gpt-4 & Assessment & Unconstrained & c & 0.376 & 0.194 & 0.533 & 0.225 & 0.000 \\
gpt-4 & Assessment & Unconstrained & e & 0.496 & 0.332 & 0.631 & 0.360 & 0.000 \\
gpt-4 & Assessment & Unconstrained & a & 0.245 & 0.051 & 0.421 & 0.082 & 0.007 \\
gpt-4 & Assessment & Unconstrained & n & 0.640 & 0.507 & 0.743 & 0.530 & 0.000 \\
gpt-4 & Acquaintance & Unconstrained & o & 0.317 & 0.119 & 0.490 & 0.152 & 0.001 \\
gpt-4 & Acquaintance & Unconstrained & c & 0.066 & -0.140 & 0.268 & -0.107 & 0.264 \\
gpt-4 & Acquaintance & Unconstrained & e & 0.114 & -0.093 & 0.312 & -0.060 & 0.139 \\
gpt-4 & Acquaintance & Unconstrained & a & 0.176 & -0.029 & 0.368 & 0.004 & 0.046 \\
gpt-4 & Acquaintance & Unconstrained & n & 0.265 & 0.063 & 0.445 & 0.096 & 0.005 \\
gpt-4 & Assistant & Unconstrained & o & 0.018 & -0.192 & 0.227 & -0.159 & 0.434 \\
gpt-4 & Assistant & Unconstrained & c & 0.062 & -0.149 & 0.268 & -0.115 & 0.282 \\
gpt-4 & Assistant & Unconstrained & e & 0.087 & -0.125 & 0.291 & -0.091 & 0.210 \\
gpt-4 & Assistant & Unconstrained & a & 0.183 & -0.028 & 0.378 & 0.006 & 0.044 \\
gpt-4 & Assistant & Unconstrained & n & 0.193 & -0.017 & 0.387 & 0.017 & 0.036 \\
\bottomrule
\end{tabular*}
\end{table}
\FloatBarrier

Correlations between self-reported and LLM-inferred personality scores across conditions and traits. Pearson's correlation coefficients, along with two-tailed 95\% confidence intervals, as well as lower-tailed 95\% confidence intervals and p-values, are reported.

\newpage

\section{User Experience Across Conditions}
\label{user_experience}
\subsection*{Group Means}
\begin{table}[h]
\small
    \begin{tabular*}{\textwidth}{@{\extracolsep{\fill}}lrrrrrr}
\toprule
& \multicolumn{2}{r}{Assessment} & \multicolumn{2}{r}{Acquaintance} & \multicolumn{2}{r}{Assistant} \\
& Conversation & Unconstrained & Conversation & Unconstrained & Conversation & Unconstrained \\
\midrule
natural & 5.240 & 5.117 & 5.631 & 5.284 & 4.642 & 4.604 \\
pleasant & 5.938 & 5.592 & 6.175 & 5.800 & 5.611 & 5.527 \\
engaging & 5.719 & 5.650 & 6.068 & 5.768 & 5.632 & 5.330 \\
questions & 5.865 & 5.864 & 6.282 & 5.800 & 3.979 & 3.714 \\
answers & 5.323 & 5.456 & 5.796 & 5.337 & 5.989 & 5.769 \\
humanlike & 4.708 & 4.835 & 5.126 & 4.842 & 4.074 & 3.857 \\
\bottomrule
\end{tabular*}
\end{table}
\FloatBarrier

\subsection*{Confidence Intervals (Lower Bounds)}
\begin{table}[h]
\small
    \begin{tabular*}{\textwidth}{@{\extracolsep{\fill}}lrrrrrr}
\toprule
& \multicolumn{2}{r}{Assessment} & \multicolumn{2}{r}{Acquaintance} & \multicolumn{2}{r}{Assistant} \\
& Conversation & Unconstrained & Conversation & Unconstrained & Conversation & Unconstrained \\
\midrule
natural & 4.937 & 4.794 & 5.397 & 4.949 & 4.318 & 4.276 \\
pleasant & 5.688 & 5.353 & 5.994 & 5.530 & 5.361 & 5.290 \\
engaging & 5.434 & 5.391 & 5.872 & 5.478 & 5.376 & 5.040 \\
questions & 5.611 & 5.624 & 6.116 & 5.508 & 3.581 & 3.353 \\
answers & 4.991 & 5.188 & 5.557 & 5.015 & 5.760 & 5.500 \\
humanlike & 4.337 & 4.491 & 4.835 & 4.456 & 3.701 & 3.466 \\
\bottomrule
\end{tabular*}
\end{table}
\FloatBarrier

\subsection*{Confidence Intervals (Upper Bounds)}
\begin{table}[h]
\small
    \begin{tabular*}{\textwidth}{@{\extracolsep{\fill}}lrrrrrr}
\toprule
 & \multicolumn{2}{r}{Assessment} & \multicolumn{2}{r}{Acquaintance} & \multicolumn{2}{r}{Assistant} \\
 & Conversation & Unconstrained & Conversation & Unconstrained & Conversation & Unconstrained \\
\midrule
natural & 5.542 & 5.440 & 5.865 & 5.620 & 4.966 & 4.933 \\
pleasant & 6.187 & 5.831 & 6.355 & 6.070 & 5.860 & 5.765 \\
engaging & 6.003 & 5.910 & 6.264 & 6.058 & 5.887 & 5.619 \\
questions & 6.119 & 6.104 & 6.447 & 6.092 & 4.376 & 4.075 \\
answers & 5.655 & 5.724 & 6.036 & 5.659 & 6.219 & 6.038 \\
humanlike & 5.080 & 5.179 & 5.417 & 5.228 & 4.446 & 4.248 \\
\bottomrule
\end{tabular*}
\end{table}
\FloatBarrier

Self-reported user experience ratings were recorded using a 7-point Likert scale. Please see SI \ref{questionnaires} for the exact wording of the items.

\newpage

\section{Demographics -  Distributions and Group Sizes}
\label{demographics}

\FloatBarrier

\begin{figure}[h]
  \centering
  \includegraphics[width=1\textwidth]{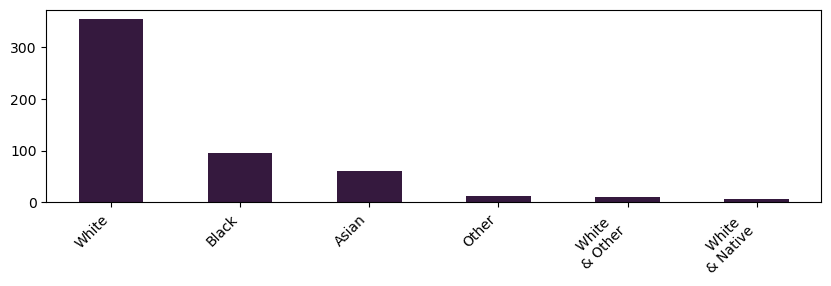}
\end{figure}

\begin{figure}[h]
  \centering
  \includegraphics[width=1\textwidth]{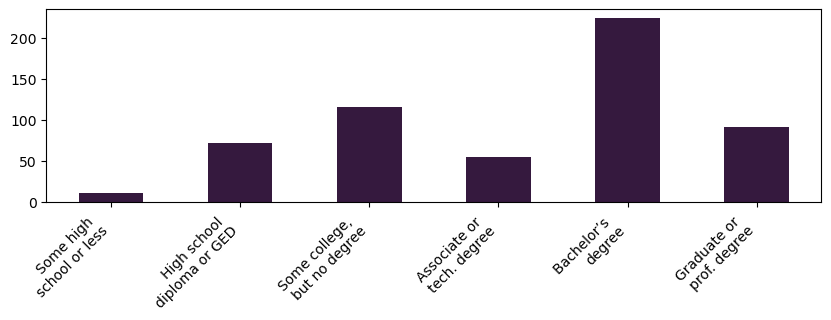}
\end{figure}

\begin{figure}[h]
  \centering
  \includegraphics[width=1\textwidth]{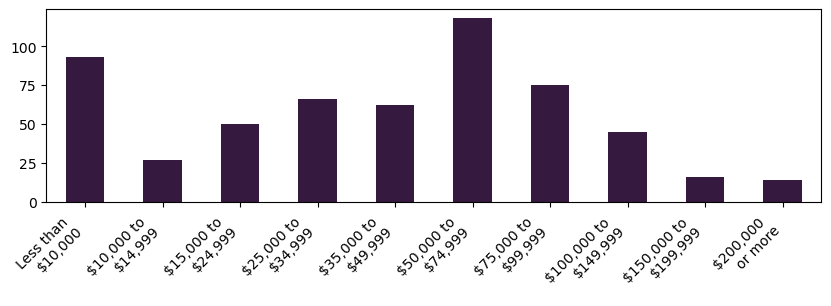}
\end{figure}

The distributions of demographic variables resulted in the following binary splits: 51.3\% female vs 48.7\% male; 51.6\% younger than 37 years of age vs 48.4 older than 37 years of age; 62.7\% identified as White vs 27.3\% for the union of all other categories; 55.7\% had at least a college education vs 44.3 with no college education; 52.7\% earned less than \$50.000 vs 47.3\% who earned more than \$50.000.

\FloatBarrier

\newpage

\section{Group Differences}
\label{group_differences}
\subsection*{Residuals}
\FloatBarrier
\begin{table}[h]
    \begin{tabular*}{\textwidth}{@{\extracolsep{\fill}}lrrrrr}
\toprule
trait & Female & Male & d & t & p \\
\midrule
o & -0.462 & -0.426 & -0.064 & -0.757 & 0.449 \\
c & -1.037 & -1.016 & -0.024 & -0.281 & 0.779 \\
e & -1.264 & -1.193 & -0.078 & -0.913 & 0.362 \\
a & -0.847 & -0.808 & -0.050 & -0.590 & 0.556 \\
n & -1.437 & -1.676 & 0.241 & 2.836 & 0.005 \\
\bottomrule
\end{tabular*}
\end{table}

\FloatBarrier
\begin{table}[h]
    \begin{tabular*}{\textwidth}{@{\extracolsep{\fill}}lrrrrr}
\toprule
trait & Younger & Older & d & t & p \\
\midrule
o & -0.432 & -0.452 & 0.037 & 0.439 & 0.661 \\
c & -0.866 & -1.175 & 0.362 & 4.303 & 0.000 \\
e & -1.103 & -1.340 & 0.261 & 3.106 & 0.002 \\
a & -0.768 & -0.885 & 0.154 & 1.828 & 0.068 \\
n & -1.745 & -1.397 & -0.351 & -4.177 & 0.000 \\
\bottomrule
\end{tabular*}
\end{table}
\FloatBarrier

\begin{table}[h]
    \begin{tabular*}{\textwidth}{@{\extracolsep{\fill}}lrrrrr}
\toprule
trait & Non-White & White & d & t & p \\
\midrule
o & -0.445 & -0.439 & -0.010 & -0.116 & 0.908 \\
c & -1.016 & -1.015 & -0.001 & -0.012 & 0.990 \\
e & -1.188 & -1.236 & 0.053 & 0.607 & 0.544 \\
a & -0.785 & -0.848 & 0.082 & 0.948 & 0.343 \\
n & -1.605 & -1.560 & -0.045 & -0.521 & 0.602 \\
\bottomrule
\end{tabular*}
\end{table}

\FloatBarrier
\begin{table}[h]
    \begin{tabular*}{\textwidth}{@{\extracolsep{\fill}}lrrrrr}
\toprule
trait & No College & College & d & t & p \\
\midrule
o & -0.499 & -0.396 & -0.185 & -2.191 & 0.029 \\
c & -1.032 & -1.002 & -0.034 & -0.407 & 0.684 \\
e & -1.295 & -1.156 & -0.153 & -1.803 & 0.072 \\
a & -0.881 & -0.779 & -0.135 & -1.592 & 0.112 \\
n & -1.586 & -1.570 & -0.016 & -0.193 & 0.847 \\
\bottomrule
\end{tabular*}
\end{table}
\FloatBarrier

\begin{table}[h]
    \begin{tabular*}{\textwidth}{@{\extracolsep{\fill}}lrrrrr}
\toprule
trait & <50K & >50K & d & t & p \\
\midrule
o & -0.495 & -0.382 & -0.203 & -2.407 & 0.016 \\
c & -0.978 & -1.057 & 0.091 & 1.078 & 0.282 \\
e & -1.169 & -1.273 & 0.114 & 1.351 & 0.177 \\
a & -0.853 & -0.793 & -0.078 & -0.927 & 0.354 \\
n & -1.693 & -1.448 & -0.245 & -2.916 & 0.004 \\
\bottomrule
\end{tabular*}
\end{table}
\FloatBarrier

Residuals were computed as the difference between inferred and self-reported personality scores, such that a negative value indicates a negative bias in inferred scores. Test statistics and p-values are reported for two-tailed t-tests.

\newpage
\subsection*{Correlations}
\begin{table}[h]
    \begin{tabular*}{\textwidth}{@{\extracolsep{\fill}}lrrrr}
\toprule
trait & Female & Male & z & p \\
\midrule
o & 0.186 & 0.330 & -1.809 & 0.070 \\
c & 0.188 & 0.185 & 0.036 & 0.971 \\
e & 0.289 & 0.291 & -0.026 & 0.980 \\
a & 0.219 & 0.261 & -0.522 & 0.602 \\
n & 0.266 & 0.370 & -1.356 & 0.175 \\
\bottomrule
\end{tabular*}
\end{table}

\FloatBarrier

\begin{table}[h]
    \begin{tabular*}{\textwidth}{@{\extracolsep{\fill}}lrrrr}
\toprule
trait & Younger & Older & z & p \\
\midrule
o & 0.250 & 0.245 & 0.063 & 0.950 \\
c & 0.195 & 0.181 & 0.172 & 0.864 \\
e & 0.307 & 0.265 & 0.541 & 0.588 \\
a & 0.234 & 0.263 & -0.366 & 0.715 \\
n & 0.305 & 0.291 & 0.182 & 0.856 \\
\bottomrule
\end{tabular*}
\end{table}
\FloatBarrier

\begin{table}[h]
    \begin{tabular*}{\textwidth}{@{\extracolsep{\fill}}lrrrr}
\toprule
trait & Non-White & White & z & p \\
\midrule
o & 0.224 & 0.259 & -0.425 & 0.671 \\
c & 0.137 & 0.211 & -0.873 & 0.383 \\
e & 0.266 & 0.291 & -0.310 & 0.757 \\
a & 0.223 & 0.265 & -0.511 & 0.610 \\
n & 0.314 & 0.296 & 0.227 & 0.820 \\
\bottomrule
\end{tabular*}
\end{table}
\FloatBarrier

\begin{table}[h]
    \begin{tabular*}{\textwidth}{@{\extracolsep{\fill}}lrrrr}
\toprule
trait & No College & College & z & p \\
\midrule
o & 0.315 & 0.161 & 1.924 & 0.054 \\
c & 0.236 & 0.124 & 1.362 & 0.173 \\
e & 0.286 & 0.280 & 0.077 & 0.939 \\
a & 0.300 & 0.193 & 1.341 & 0.180 \\
n & 0.336 & 0.258 & 1.006 & 0.314 \\
\bottomrule
\end{tabular*}
\end{table}
\FloatBarrier

\begin{table}[h]
    \begin{tabular*}{\textwidth}{@{\extracolsep{\fill}}lrrrr}
\toprule
trait & <50K & >50K & z & p \\
\midrule
o & 0.296 & 0.193 & 1.296 & 0.195 \\
c & 0.242 & 0.056 & 2.255 & 0.024 \\
e & 0.230 & 0.311 & -1.033 & 0.301 \\
a & 0.321 & 0.135 & 2.327 & 0.020 \\
n & 0.293 & 0.261 & 0.410 & 0.682 \\
\bottomrule
\end{tabular*}
\end{table}
\FloatBarrier

Correlations were computed as Pearson’s correlation coefficients between inferred and self-reported personality scores within each group. A larger value indicates higher accuracy for the specified group. Test statistics and p-values are reported for Fisher's z-test for differences in correlation coefficients.

\end{document}